\documentclass[aps,pra,twocolumn,groupedaddress]{revtex4}
\usepackage{amsmath}
\usepackage{amssymb}
\usepackage{latexsym}
\usepackage{bm}
\usepackage{graphicx}
\usepackage{dcolumn}
\usepackage{tipa}
\usepackage{txfonts}
\usepackage[caption=false]{subfig} 
\usepackage{changes}

\usepackage[colorlinks=true,linkcolor=blue,citecolor=blue,urlcolor=blue]{hyperref}

\begin{document}

\title{A High-Contrast Bragg Atom Interferometer for Testing Continuous Spontaneous Localization}

\author{Huaiyu Zhu$^{1}$}
\author{Ju Liu$^{1}$}
\author{Tao Zhang$^{1}$}
\author{Qin Luo$^{1}$}\email[E-mail: ]{luoqin@hust.edu.cn}
\author{Zhongkun Hu$^{1,2}$}
\author{Minkang Zhou$^{1}$}\email[E-mail: ]{zmk@hust.edu.cn}

\affiliation{$^1$National Gravitation Laboratory, MOE Key Laboratory of Fundamental Physical Quantities Measurement, and School of Physics, Huazhong University of Science and Technology, Wuhan 430074, People's Republic of China\\
$^2$Wuhan Institute of Quantum Technology, Wuhan 430206, People’s Republic of China} 

\begin{abstract}
The continuous spontaneous localization (CSL) model is one of the most promising approaches to address the wave function collapse problem in the measurement process of standard quantum mechanics. In this work, the effect of the CSL model on a Bragg atom interferometer was investigated. A Bragg interferometer achieving high fringe contrast of 99$\%$ has been demonstrated, maintaining this performance level at interrogation time up to $T$ = 60 ms. The primary factors responsible for fringe contrast loss in the atom interferometer were systematically analyzed and corrected. This improvement established a new upper limit of $\lambda_{CSL}=1.27\times10^{-5}$ s$^{-1} @ r_C=10^{-5}$ m for the CSL collapse rate, representing approximately 4 times enhancement over previous atom-interferometric constraints.
\end{abstract}

\maketitle
\section{introduction}
Quantum mechanics is undoubtedly one of the most successful theories in modern physics, with its predictions confirmed by countless experiments and observations. 
However, despite its remarkable empirical success, standard quantum mechanics does not solve the measurement problem - quantum system undergoing deterministic Schrödinger evolution is then followed by a probabilistic evolution and yields a  random result \cite{red2,red3,D}. 
Consequently, the collapse model was proposed as a more universal phenomenological dynamical model.

These collapse models are typically achieved by adding terms to the linear Schrödinger equation which describes the collapse of the particle's wave function in position space to a localized state\cite{A1, pearle1976, diosi1989, A4, A2}. In these collapse models, the continuous spontaneous localization(CSL) model is most studied \cite{piccione2025exploring}, and it describes the nonlinear and stochastic terms of collapse through two newly introduced monomial parameters, $\lambda_{\mathrm{CSL}}$ and $r_{C}$ \cite{A4}, where $\lambda_{\mathrm{CSL}}$ represents the collapse rate and determines the collapse's intensity, while $r_{C}$ represents the collapse's spatial resolution.

The bound of the two parameters can be determined through various experiments, thereby testing the CSL model. Experimental tests of the CSL model can divide into non‑interferometric tests \cite{B36,B38,B40,B42,B43,B44,N1,N2} and interferometric tests \cite{B26,B27,A9,pedalino2026probing,A,B28}, each type examining the model from distinct perspectives.
Interferometric tests are the most direct approach for probing the CSL model\cite{red2022present}, but they face difficulties from a technical perspective, particularly in preparing and maintaining spatial superpositions of massive systems over time.
The most massive system for interferometric tests of the CSL model currently employs a Talbot-Lau interferometer using sodium nanoparticles with a mass of approximately $172,000$ atomic mass units, giving a bound of $\lambda_{\text{CSL}} \approx 10^{-8}\text{s}^{-1}$ at $r_C = 10^{-7}\text{m}$ \cite{pedalino2026probing}, but with relatively weak bound in the region above $10^{-7}\text{m}$.

As an important tool for high-precision measurement, atom interferometers are widely used in the measurement of gravity \cite{Robins2011cold,C3,Bouyer2018gravity} and its gradients \cite{zhou2010precisely, stray2022quantum}, determining the gravitational constant \cite{tino2008determination, tino2014precision}, probing dark energy \cite{burrage2015probing}, searching spin-dependent exotic interactions\cite{shu2024constraint} and testing the CSL model \cite{A}.
For interferometric tests of the CSL model, atom interferometers offer the distinct advantages of larger spatial separations, longer coherence times, and robust controllability, and place stringent constraints on the collapse parameters at large values of $r_C$.
However, atom interferometry measurements often exhibit limited precision, as current analyses typically ascribe the entire decay of fringe contrast with increasing pulse interval time $T$ to CSL effect \cite{A}. This approach inadequately accounts for contrast loss from other physical factors, thereby compromising the precision of such tests.

In this work, we perform a test of the CSL model utilizing a Bragg atom interferometer with high fringe contrast. It is worth noting that in the framework of standard quantum mechanics, the decay of interference signals due to initial state uncertainties can be rigorously modeled using the "least-biased description" based on the maximum entropy principle \cite{englert1994least, zaugg1994theory}. While these theoretical models provide a fundamental understanding of dephasing mechanisms, our work emphasizes a direct experimental analysis. In this atom interferometer, factors contributing to the decay of fringe contrast with increasing interrogation time $T$—such as the misalignment, transverse diffusion, and longitudinal velocity spread of atomic cloud—have been systematically characterized and mitigated. This comprehensive approach has improved the precision of the upper bound on the CSL collapse rate $\lambda_{\mathrm{CSL}}$ by nearly 4 times at $r_{C}=10^{-5}$ m compared to previous such tests.
\vspace*{\fill} %
\section{PRINCIPLES}
\subsection{Bragg Atom Interferometer}\label{sec:IIA}
\begin{figure}[h]
  \centering
  \includegraphics[width=\linewidth]{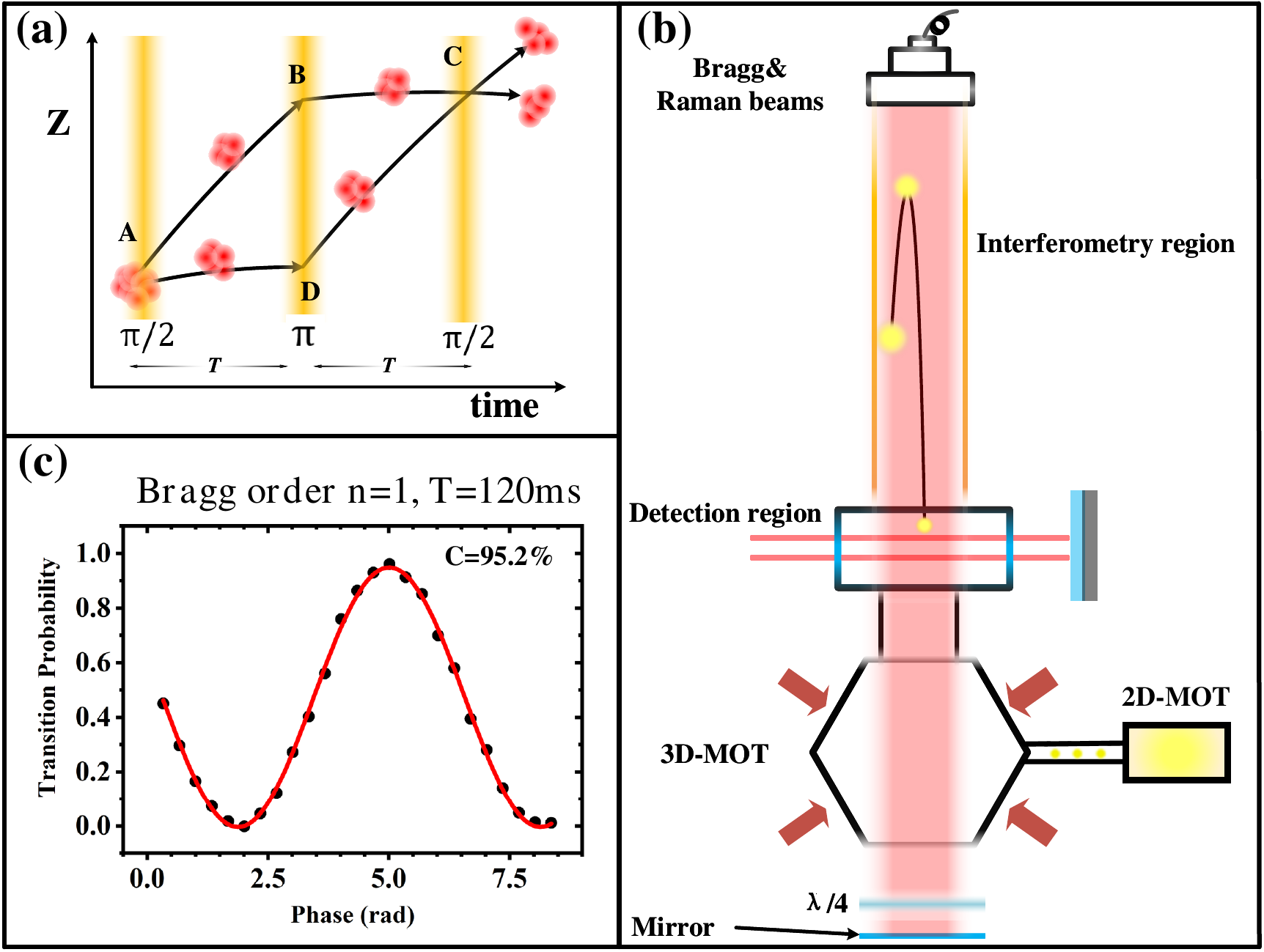} 

  \caption{(a) The space–time diagram of Mach–Zehnder–type atom interferometer.
  (b) Schematic of our laser system (see the main text for the experimental sequence). The atomic cloud continuously diffuses along its path, deviating from the center of the pulsed light.
  (c) Interference fringes at $n=1$, $T=120$ ms, deriving the interference fringe formula and fringe contrast through sine fitting}.
  \label{fig:1}

  \refstepcounter{subfigure}\label{fig:1a}
  \refstepcounter{subfigure}\label{fig:1b}
  \refstepcounter{subfigure}\label{fig:1c}
\end{figure}

In a Mach–Zehnder configuration atom interferometer, the atomic wave packet is split, reflected, and recombined via a sequence of Bragg pulses $\pi/2$–$\pi$–$\pi/2$, forming interference, as shown in Fig.~\ref{fig:1a}.
Atoms initially in the momentum state $|p_0\rangle$ interact with a $\pi/2$ Bragg pulse at point A, being coherently split into two momentum components. One part retains its original momentum, corresponding to velocity $v_1$, while the other part acquires an additional momentum of $n\hbar k_{\mathrm{eff}}$, attaining velocity $v_2$,
where $k_{\mathrm{eff}} = k_1 - k_2$ is the effective wave vector and $n$ denotes the Bragg diffraction order. After a time interval $T$, atoms encounter a $\pi$ Bragg pulse. This pulse reflects the two wave packets, reducing the momentum of the upper path by $n\hbar k_{\mathrm{eff}}$ and increasing the momentum of the lower path by the same amount. After another time interval $T$, atoms are subjected to the second $\pi/2$ Bragg pulse, causing the two paths to recombine and interfere. Following this Mach--Zehnder type sequence, the probability for an atom to be found in the initial momentum state $|p_0\rangle$ is 
\begin{equation}
P = \frac{1}{2} \left[1 + C\cos(\Delta\phi)\right],
 \label{eq1}
\end{equation}
where $C$ is the fringe contrast of the interferometer, $\Delta\phi$ represents the accumulated phase difference between the two interferometric paths. By linearly scanning the phase, interference fringes can be obtained, from which both the fringe contrast and initial phase are extracted, just like Fig.\ref{fig:1c}.
\subsection{Test Continuous Spontaneous Localization}\label{sec:IIB}
According to the calculations presented in Reference \cite{A}, the CSL model does not induce phase shifts in atom interferometers but rather affects the interference fringe contrast. Accordingly, Eq. (1) can be reformulated as follows
\begin{equation}
P = \frac{1}{2} \left[ 1 + \exp\left( -\frac{E[\Delta\phi_{\mathrm{CSL}}^2]}{2} \right) \cos(\Delta\phi) \right]
\label{eq2}
\end{equation}
where $\Delta\phi_{\mathrm{CSL}}$ is the phase noise introduced by the CSL model. In Eq.~(2), the second-order moment $E[\Delta\phi_{\mathrm{CSL}}^2]$ is
\begin{equation}
E\left[ \Delta \phi_{\mathrm{CSL}}^2 \right] = 4 \lambda_{\mathrm{CSL}} N^2 T 
\left[ 1 - \frac{\sqrt{\pi}}{2}  \frac{ \mathrm{erf} \left( \frac{v_2 - v_1}{2 r_C} T\right) }{ \frac{v_2 - v_1}{2 r_C}T } \right]
\label{eq3}
\end{equation}
where \( v_1 \) and \( v_2 \) represent the velocities of atoms along the two interferometer arms, as described previously and illustrated in Fig.~\ref{fig:1a}. The symbol \( N \) denotes the number of \(^{87}\mathrm{Rb} \) nucleus, and $T$ is the time interval between two Bragg pulses.

By substituting Eq.~\eqref{eq3} into Eq.~\eqref{eq2}, the complete expression for the interference fringe under the influence of the CSL effect can be obtain. Based on the definition of fringe contrast, the following relationship can be derived
\begin{equation}
\ln C(T) = \ln C_0 - 2 \lambda_{\mathrm{CSL}} N^2 T \left[ 1 - \frac{\sqrt{\pi}}{2}  \frac{ \mathrm{erf} \left( \frac{v_2 - v_1}{2 r_C} T \right) }{ \frac{v_2 - v_1}{2 r_C} T } \right].
\label{eq4}
\end{equation}

The result shows that the contrast decays exponentially, and the decay rate depends on the collapse rate $\lambda_{CSL}$, the correlation length $r_C$, the number of  nucleons $N^2$, the interval time $T$, and the velocities of the two wave packets $v_1$ and $v_2$.

In Eq.~\eqref{eq4}, $\mathrm{erf}$ represents the error function, whose independent variable is $\frac{v_2 - v_1}{2r_C}T$. Depending on the value of $r_C$, two distinct correlation regimes exist:
\begin{itemize}
    \item[(i)] When \( r_c \gg (v_2 - v_1)T \approx 10^{-3} \, \text{m} \), $r_C$ is much larger than the maximum separation between the two arms of the interferometer, the functional dependence on $\lambda_{CSL}$, $r_C$, $C$ and $T$ remains unchanged. As $r_C$ increases, the contrast loss becomes smaller. As $r_C$ increases, the contrast loss decreases.  Therefore, the bound on $\lambda_{CSL}$ weaken as $r_C$ increases.
\end{itemize}
\begin{itemize}
    \item[(ii)] When \( r_C \ll (v_2 - v_1)T \), the error function component in Eq.~\eqref{eq4},  \(\mathrm{erf}\left( \frac{v_2 - v_1}{2r_C}T \right) / \left( \frac{v_2 - v_1}{2r_C}T \right) \), becomes negligible. As a result, the  damping factor \( E[\Delta\phi^2_{CSL}] \) becomes independent on \( r_C \). Consequently, the influence of $r_C$ on the interference contrast vanishes. Therefore, the variation of $\lambda_{CSL}$ is independent on $r_C$, and Eq.~\eqref{eq4} can be simplified to
    \begin{equation}
     \ln C(T) = \ln C_0 - 2\lambda_{CSL} N^2 T.
     \label{eq5}
    \end{equation}
\end{itemize}

Furthermore, the influence of the size of atomic wave
packet need to be considered separately when $r_C$ is very small \cite{A}. As $r_c$ becomes smaller, the localization step length increases, resulting in a larger position variance induced by the localization effect. To observe interference, the two wave packets in the interferometer must exhibit sufficient spatial overlap at time 2$T$. Conservatively, the position variance induced by the CSL effect needs to be smaller than one tenth of the wave-packet size at time 2$T$. This condition leads to an upper bound on the ratio $\lambda_{\text{CSL}} / r_C^2$. We take the initial wave packet size $\sigma$=$0.5\times10^{-7}$ m here. The bound is computed as
\begin{equation}
\frac{\lambda_{\text{CSL}}}{r_C^2} \leq 1.6 \times 10^7 \, \text{m}^{-2} \, \text{s}^{-1}.
\label{eq6}
\end{equation}
It gives a stronger bound for CSL parameters than Eq.~\eqref{eq5}.
\section{EXPERIMENT Setup}
To test the CSL model, the fringe contrasts at different interval $T$ in a Bragg atom interferometer are measured. A schematic of the experimental apparatus is shown in Fig.\ref{fig:1b}. About $10^9$ atoms are loaded in a three-dimensional magneto-optical trap (3D-MOT) within 200 ms with the help of a two-dimensional magneto-optical trap. Then, atoms are launched by modifying the frequency of upward trapping lasers, and cooled during the moving molasses, where the atomic cloud reaches temperature of 7 $\mu$K. After being launched, the atomic cloud enters the detection region for Raman sideband cooling and longitudinal velocity selection.

When the atomic cloud reaches the detection region, Raman sideband cooling (RSC) are applied during their motion through the region to lower the temperature in the horizontal directions. RSC can be completed within a few milliseconds, much shorter than the time required for evaporative cooling, and can retain a large number of atoms. The RSC of cold atoms has been described in \cite{C49}. After completing the RSC, the atom's temperature in the horizontal directions is approximately 0.85 $\mathrm{\mu K}$\cite{C}.

After applying RSC to bring the atoms into the $\left|5^2 S_{1/2}, F=1\right\rangle$ state, most atoms are in a magnetically sensitive state $\left|5^2 S_{1/2}, F=1, m_F=1\right\rangle$. To perform the experiment, the atoms need to be prepared in magnetically insensitive state $\left|5^2 S_{1/2}, F=1, m_F=0\right\rangle$. A pump laser with frequency $F=1\rightarrow F'=0$ is applied to drive the transition $\left|F=1, m_F=\pm1\right\rangle \to \left|F'=0, m_F=0\right\rangle$. All atoms spontaneously transition back to the $\left|F=1, m_F=0\right\rangle$ state. Then, a microwave pulse pumps all atoms into the $\left|F=2, m_F=0\right\rangle$ state. Subsequently, an $F=1$ blowaway laser removes any atoms that remained in the $F=1$ state. Following this, a velocity-sensitive Raman laser is applied along the longitudinal direction to prepare atoms in the $\left|F=1, m_F=0\right\rangle$ state with a narrow velocity distribution. An $F=2$ blowaway laser is then used to remove any remaining atoms in the $F=2$ state. This procedure yields an ultracold atomic ensemble with a longitudinal velocity distribution of only $0.08v_r$\cite{C}.
 
As shown in Fig.\ref{fig:1b}, after initial state preparation, the center of the atomic cloud is essentially coincident with the center of the pulse beam, and the cloud remains tightly compact. 
Then, a Mach-Zehnder interferometer is constructed by applying a Bragg pulse sequence of $\pi/2 - \pi - \pi/2$ when the atom cloud enter the interference region. 
During this sequence, the atomic cloud continuously diffuses, and owing to its transverse velocity, it steadily drifts from the center toward the periphery. In this experiment, the first-order Bragg diffraction $n$ = 1 is employed, and the pulse width of the $\pi$ pulse is 62 $\mathrm{\mu s}$, with an efficiency of 99$\%$. 

The method of detecting the final atomic states after interference is based on Raman spectroscopy detection \cite{C39}. This allows us to perform Raman scattering to selectively detect atoms in different momentum states,  selecting atoms in distinct momentum states via the Raman pulse and detecting them separately to obtain interference fringes. The interference fringe at $T$=120 ms is shown in Fig.\ref{fig:1c}. 
The whole experimental procedure can be found in \cite{C, C4, C43}.

By modulating the free evolution time $T$, the central fringe can be obtained, where the interference fringe contrast reaches its maximum value. The interference fringe contrasts at different $T$ is shown in Table \ref{tab1}.
\begin{table}[!h]
\caption{Fringe contrasts at different interference times $T$.}
\footnotesize 
\centering
\resizebox{0.48\textwidth}{!}{
\renewcommand{\arraystretch}{1.8} 
\begin{tabular}{@{\hspace{10pt}}c@{\hspace{10pt}}c@{\hspace{10pt}}c@{\hspace{10pt}}c@{\hspace{10pt}}c@{\hspace{10pt}}c@{\hspace{10pt}}c@{\hspace{10pt}}c@{\hspace{10pt}}c@{\hspace{10pt}}}
\hline\hline
$T$ (ms)  & 10 & 60 & 120 & 150 & 180 & 210 & 230 & 250 \\
\hline
$C$ ($\%$)  & 99.0 & 99.0 & 95.2 & 80.0 & 72.9 & 59.8 & 58.8 & 49.7 \\
\hline
Uncertainty($\%$) & 1.7 & 1.8 & 0.9 & 2.3& 2.5 & 2.6 & 2.6 & 2.9 \\
\hline\hline
\end{tabular}
}
\label{tab1}
\end{table}

As shown in Table \ref{tab1}, the high fringe contrast and relatively small uncertainty of fringe contrast we obtained depends on the effective suppression of vibration-induced noise, beyond the state preparation and velocity selection of the atomic cloud. In a light-pulse atom interferometer, the vibration of the retroreflection mirror is directly transferred to the interferometer phase through the Bragg pulses, and the corresponding acceleration-induced phase noise scales approximately as $k_{\text{eff}}aT^{2}$. Therefore, vibration control becomes particularly crucial at extended pulse separation times. 
The experimental apparatus used in this work \cite{C} is equipped with an active vibration isolation platform developed by our group \cite{C43}. Evaluations of this platform demonstrate that even at a long pulse separation time of $T = 320$~ms, the residual vibration noise contribution is strongly suppressed to approximately $0.7\,\mu\text{Gal}/\text{Hz}^{1/2}$. 
This effective suppression of vibration noise fundamentally guarantees the high-quality fringe obtained in our experiment.

\section{Result of testing CSL}
As shown in Table \ref{tab1}, the fringe contrast still remains 49.7$\%$ at $T = 250$ ms. However, the interference fringe contrast decreases by $1\%$ to $50\%$ within the range of $T = 10$ ms $\sim$ $250$ ms. If all fringe contrast loss is attributed to the CSL effect, the impact of CSL would be overestimated, implying a weaker bound on the CSL parameters. Therefore, to establish more stringent constraints on the CSL parameters, it is essential to identify and account for the dominant factors responsible for the degradation of atom interferometric fringe contrast with increasing interrogation time $T$ in the experimental results.  

In atom interferometers, the primary sources of fringe contrast degradation typically include the position misalignment and transverse velocity of the atomic cloud, as well as its transverse and longitudinal diffusion. In our experimental setup, the Bragg laser beam possesses a spatial Gaussian profile with a transverse beam waist of $w_0 = 10\text{ mm}$\cite{C}. Due to the spatial inhomogeneity arising from this Gaussian distribution, the local laser intensity varies significantly across the transverse plane. Consequently, as the atomic cloud undergoes transverse diffusion or deviates from the beam center due to initial misalignment and transverse velocity, atoms at different positions experience varying local laser intensities. This spatial dependence introduces inhomogeneities in the atom-light interaction across the atomic ensemble, leading to imperfect diffraction efficiencies during the pulse sequence, which acts as the fundamental physical mechanism driving the significant loss of fringe contrast.

\begin{figure}[b]
    \centering
    \includegraphics[width=8cm]{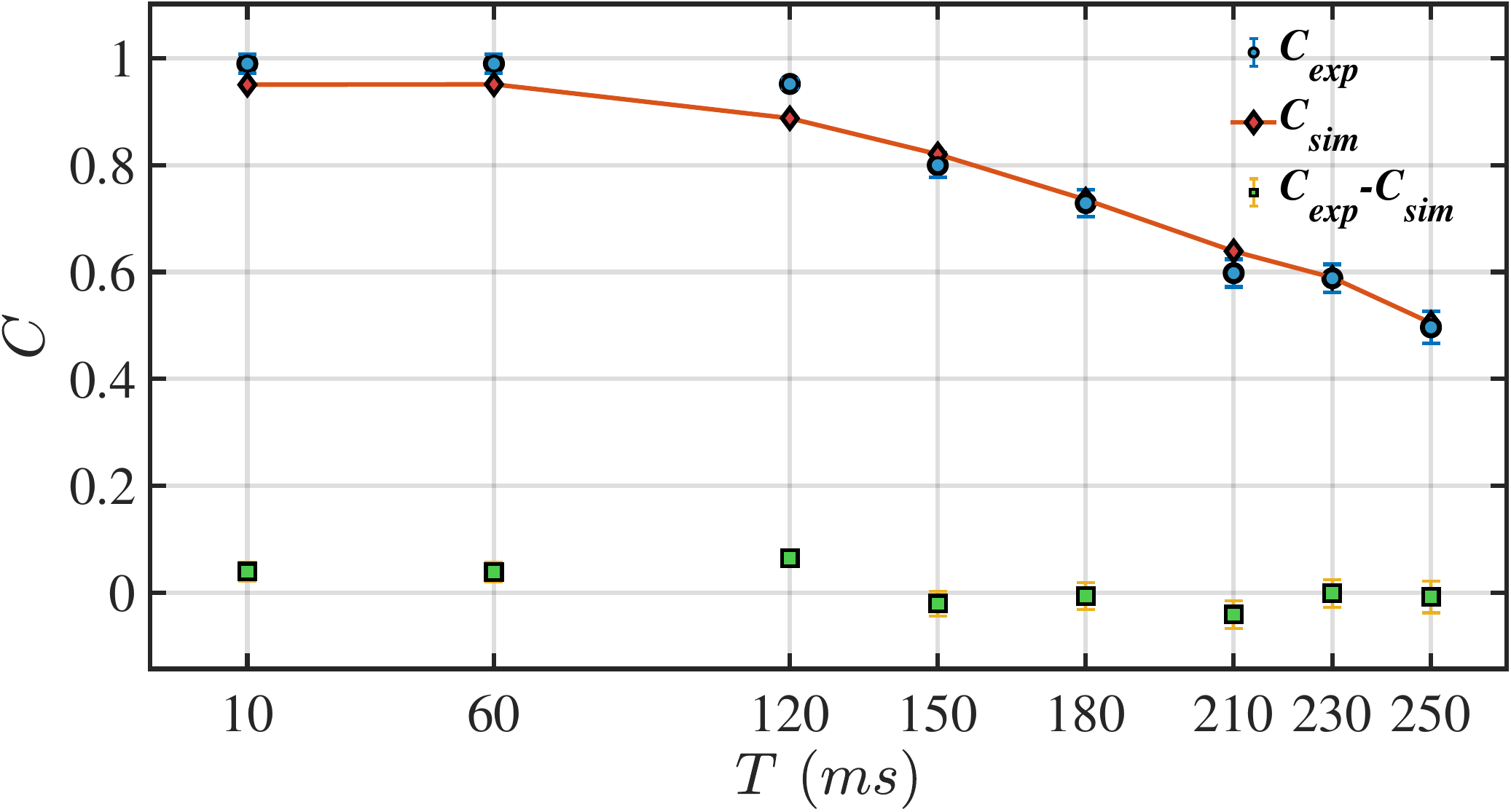}
    \caption{The simulated fringe contrast obtained from the Monte Carlo compared with the experimental results by the Bragg atom interferometer. $C_{\text{exp}}$ represents the fringe contrast obtained from our actual experiments, while $C_{\text{sim}}$ denotes the contrast derived from simulation.}
    \label{fig5}  
\end{figure}

The fringe contrast in atom interferometers is commonly simulated using the well-established Monte Carlo method, as detailed in Appendix~\ref{APPA}. The simulation results is shown by the red diamonds in Fig.~\ref{fig5}, while the squares represent the experimentally measured fringe contrasts. This simulation are based on experimental parameters that have been precisely measured, including the cloud's initial transverse position, horizontal velocity, transverse and longitudinal temperature, all listed in Table~\ref{tab2}.

\begin{table}[htbp]
\caption{Atomic cloud parameters after velocity selection.}
\footnotesize 
\centering
\resizebox{0.48\textwidth}{!}{
\renewcommand{\arraystretch}{1.8} 
\begin{tabular}{@{\hspace{10pt}}c@{\hspace{10pt}}c}
\hline\hline
\textbf{Atomic cloud Parameters} & \textbf{Measured Value} \\
\hline
horizontal position width in NS (mm) & $0.8 \pm 0.5$ \\
\hline
horizontal position width in EW (mm) & $0.8 \pm 0.5$ \\
\hline
horizontal moving velocity in NS (mm/s) & $1.4 \pm 0.7$ \\
\hline
horizontal moving velocity in EW (mm/s) & $5.2 \pm 0.6$ \\
\hline\hline
\end{tabular}
}
\label{tab2}
\end{table}

To rigorously test the CSL model, it is necessary to subtract the contrast loss caused by other factors. After obtaining the results of the Monte Carlo simulation, the fringe contrast can be optimized accordingly. In the simulation, the contrast reduction originates solely from the atomic ensemble’s transverse position and mean velocity, longitudinal velocity spread, and transverse velocity spread. This simulated contrast loss is denoted as $\Delta = 1 - C_{sim}$. 

In the actual experiment, fringe contrast loss arises not only from expansion and misalignment, but also from the potential CSL effect. Therefore, the true contrast including CSL effects is given by $C'_{\text{exp}} = C_{\text{exp}} + \Delta$, where the observed contrast reduction with respect to the interferometer time $2T$ can then be attributed purely to the CSL mechanism.

\begin{figure}[h]
    \centering
    \includegraphics[width=8cm]{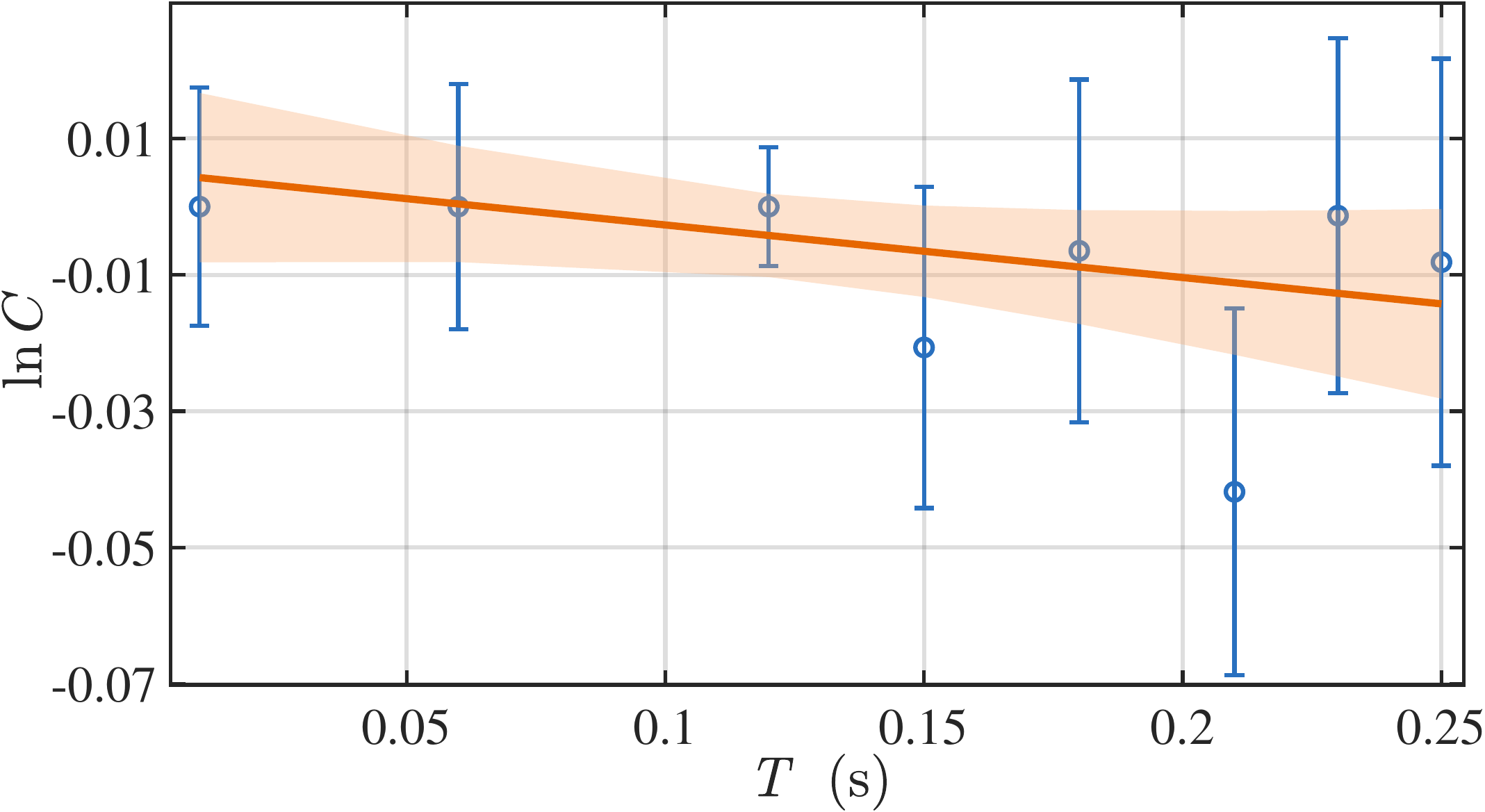}
    \caption{The corrected fringe contrast $C'_{\text{exp}}$ as a function of of the Bragg pulse separation time $\text{T}$. Error bars denote the $1\sigma$ uncertainty in the fitting contrast. The solid line is a weighted least-squares fit of Eq.~\eqref{eq5}, and the orange shaded region represents the $1\sigma$ confidence interval of the weighted least-squares fit.}
    \label{fig6}  
\end{figure}
Figure~\ref{fig6} shows the dependence of the corrected fringe contrast $C'_{\text{exp}}$ on the total interferometer time $T$ within the range of 0 ms to 250 ms. The solid line represents a weighted least-squares fit to the contrast decay function ln$C(T)$ based on Eq.~\eqref{eq5}. From this line, the CSL collapse rate of $\lambda_{\text{CSL}} \leq (5.1 \pm 6.4) \times 10^{-6}\,\mathrm{s}^{-1}$ is derived.
\begin{figure}[htbp]
    \centering
    \includegraphics[width=8.5cm]{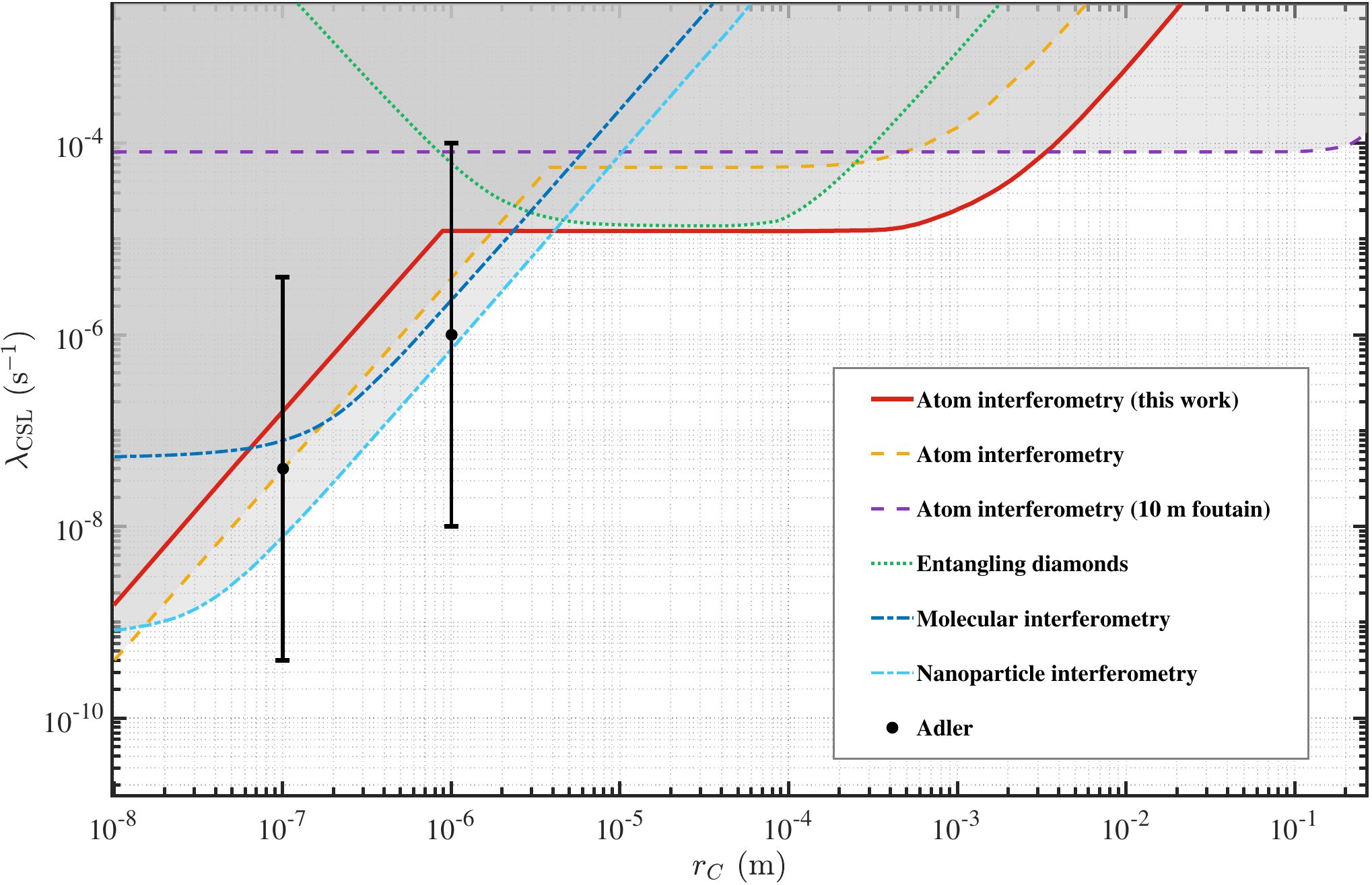}
    \caption{Exclusion plot for the CSL parameters comparing different interferometric experiments: 
    The red solid line illustrates our result.
    The yellow dashed line\cite{A} and purple dashed line\cite{A33} correspond to previous results from other atom-interferometry experiments.
    The green dotted line denotes the bounds from entangled diamonds interferometry\cite{A34,A35}. 
    The dark blue dash-dot line shows the molecular interferometry experiments\cite{A9,A36,A37}, while the cyan dash-dot line denotes the bounds from nanoparticle interferometry\cite{pedalino2026probing}. 
    Shaded areas indicate the excluded regions and the black dots is Adler’s theoretical predictions\cite{A16}.  
    }
    \label{fig7}  
\end{figure}

Furthermore, the uncertainties contributed by the parameters of the atomic cloud (listed in Table~\ref{tab2}) should be considered in the Monte Carlo simulation. These collectively contribute an additional uncertainties to bound on $\lambda_{\text{CSL}}$, is $4.1 \times 10^{-6}\,\mathrm{s}^{-1}$. Taking this into account, and based on the approximate model described in Eq.~\eqref{eq5}, the $\lambda_{\text{CSL}} \leq (5.1 \pm 7.6) \times 10^{-6}\,\mathrm{s}^{-1}$ is obtained. It should be noted that the actual uncertainty is less than $7.6 \times 10^{-6}\,\mathrm{s}^{-1}$. This is because the uncertainty obtained from the weighted least squares fitting actually originates from the contrast error during fringe processing. Since the uncertainty of the fringe contrast may share the same source as the uncertainty contributions listed in Table~\ref{tab2}, leading to an overestimation of the collapse rate uncertainty. Thus, by conservatively adding the maximum uncertainty to the central value, We give the limit of $\lambda_{\text{CSL}} \leq1.27\times10^{-5}$ s$^{-1}$ as the final collapse rate constraint for greater rigor. 

According to the discussion in principle section, the exclusion of $\lambda_{\mathrm{CSL}}$ and $r_{C}$ is shown as red line in Fig.~\ref{fig7}, and other results from other interferometric experiments are also displayed in the figure. In Fig.~\ref{fig7}, for larger values of $r_{C}$, the parameter $\lambda_{\mathrm{CSL}}$ can be obtained directly by substituting $C'_{\text{exp}}$ into Eq.~\eqref{eq3}. Over the range of $10^{-6}$ m<$r_C$< $3\times10^{-4}$ m, the value of $\lambda_{\mathrm{CSL}}$ remains nearly constant. As determined from the preceding calculation, it is $1.27\times10^{-5}$ s$^{-1}$. When $r_c$ is smaller, stronger bounds can be derived from Eq.~\eqref{eq6}, the value of $\lambda_{\text{CSL}.}/r_c^2 \leq 4.4 \times 10^6 \, \text{m}^{-2}$. 

\section{Conclusion and Discussion}
In conclusion, we have developed a high-contrast Bragg atom interferometer and systematically identified the dominant factors responsible for fringe contrast degradation in this apparatus. 
After accounting for these effects, the experimental constraint on the collapse rate parameter $\lambda_{CSL}$ in the continuous spontaneous localization model has been improved by nearly a factor of four, reaching $1.27\times10^{-5} s^{-1}$ at a characteristic radius $r_C=10^{-5}$ m. It is important to note that our results do not constitute a definitive proof of the CSL model, but rather establish rigorous upper bounds on its parameters by attributing any unmodeled contrast loss, which may include unknown characteristics of the detection device—to the collapse mechanism.

To date, the limitation in testing the CSL model using atom interferometry primarily arises from non-CSL sources of contrast loss. 
Due to this contrast-loss mechanism, further research can focus on suppressing the contrast loss, such as using quantum optimal control\cite{QOC}; or employing experimental setups that extend the interferometry times, such as long-baseline atomic interferometers\cite{A33, zhou2015test} and optical lattices\cite{lattice2024}.
Furthermore, based on the standard CSL model, a broader framework for Poissonian spontaneous localization(PSL) model have proposed\cite{2025}, which also provides implications for the development of experimental tests.

\section*{ACKNOWLEDGMENTS}
We thank Professor Sandro Donadi, Professor Yiqiu Ma, Professor Xiaochun Duan and Qi Dai for enlightening discussions. This work is supported by the National Natural Science Foundation of China (Grant Nos. U2541241, U2341247, 12374464) and National Key Research and Development Program of China (Grant No. 2023YFC2907003, 2022YFC3003802).

\appendix  
\section{Monte Carlo simulation}\label{APPA}  

\begin{figure}[htbp]
    \centering
    \includegraphics[width=8cm]{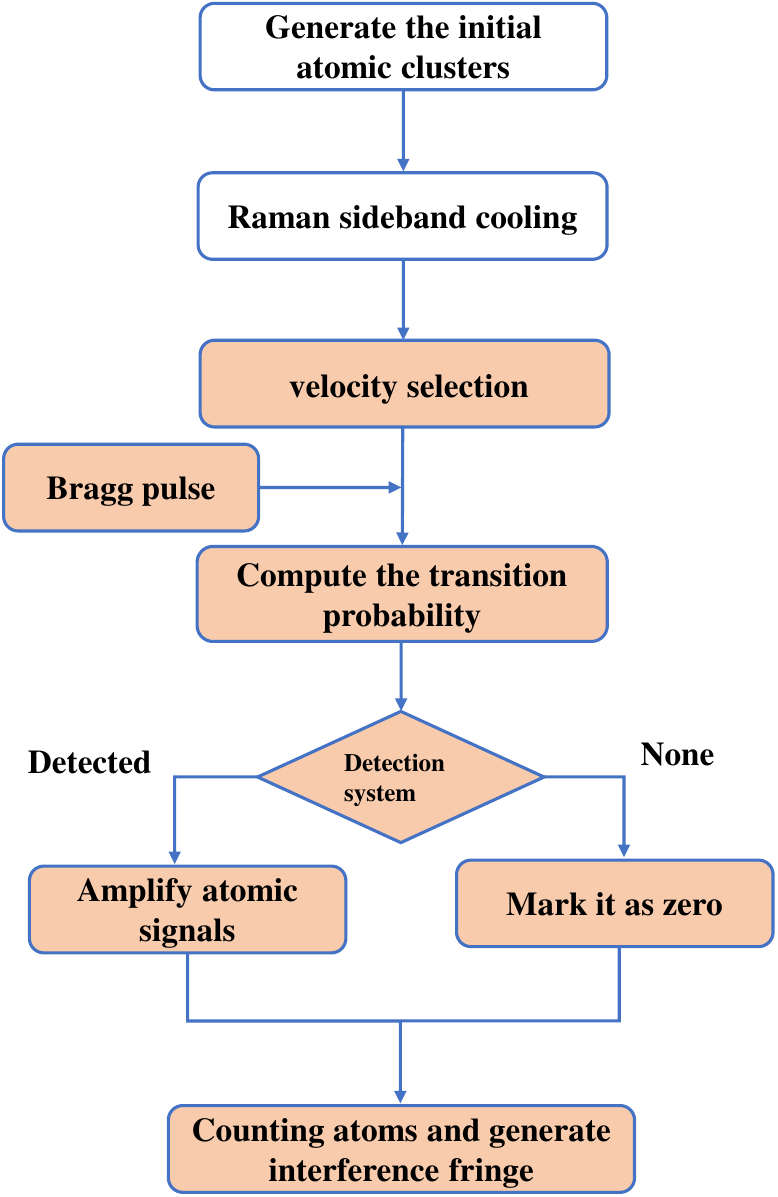}
    \caption{Simulate the flowchart according to the actual experiment. }
    \label{fig4}  
\end{figure}

We employ Monte Carlo simulation for analysis the fringe contrast loss in the atom interferometer. The complete simulation workflow is shown in Figure~\ref{fig4}.

In order to minimize the potential errors that might arise during the simulation process and to maintain consistency with the experiment, we did not generate initial atomic cloud in 3D-MOT and then perform RSC and velocity selection simulation on it; instead, we generated atomic cloud based on the measurement results after velocity selection. In practice, our fitting process commences with the light brown fill step in Figure~\ref{fig4}. 

After velocity selection, the $\pi$-pulse efficiency reaches about $99\%$, so that, in the Bragg diffraction process implemented in our experiment, the multilevel atomic system can be effectively reduced to a two-level system. For such a two-level system, the Hamiltonian can be written as
\begin{equation}
H = \frac{1}{2}
\begin{pmatrix}
\delta(t) & \Omega(t) \\
\Omega(t) & -\delta(t)
\end{pmatrix}
= \frac{1}{2}\bigl[\delta(t)\,\sigma_z + \Omega(t)\,\sigma_x\bigr],
\end{equation}
where $\delta(t)$ is the detuning between the laser and the atomic transition, $\Omega(t)$ is the Rabi frequency during the interferometric sequence, and $\sigma$ denotes the Pauli matrices.

Due to the inhomogeneous intensity profile of the Gaussian beam and the finite spatial extent of the atomic cloud, the effective Rabi frequency is position dependent and can be written as 
\begin{equation}
\Omega(r,t) = \Omega_0(t)\exp\left(-\frac{r^2}{w^2}\right),
\label{A2}
\end{equation}
here $w$ is the $1/e^2$ radius of the Gaussian beam, and $r$ is the distance of the atom from the center of the beam.

Here, the position distribution of the atomic cloud in the transverse direction is
\begin{equation}
f(\mathbf r) = f(r_x,r_y)
= \frac{1}{2\pi\sigma_{r_x}\sigma_{r_y}}
\exp\left[
-\frac{(r_x-\bar r_x)^2}{2\sigma_{r_x}^2}
-\frac{(r_y-\bar r_y)^2}{2\sigma_{r_y}^2}
\right],
\end{equation}
where $r_x$ and $r_y$ denote the two horizontal coordinates and $\sigma_{r_x}$, $\sigma_{r_y}$ are the corresponding position spread. We choose the x axis to point north–south and the y axis east–west. The velocity distribution in the transverse directions is also Gaussian, with an analogous form.  

For the longitudinal direction, we have only considered the longitudinal velocity distribution and have not taken the influence of positional distribution on into account. This is because its effect is negligible, and moreover, after velocity selection, the temperature of atomic cloud in the longitudinal direction is extremely low, making the impact of diffusion virtually negligible.

Furthermore,  to achieve higher diffraction efficiency within a shorter time, and thus suppress higher-order Bragg diffraction, we choose a Gaussian temporal envelope for the pulse\cite{yyk}. During free evolution we set $\Omega_0(t)=0$,
while during the atoms-pulse interaction it is given by: 
\begin{equation}
\bar{\Omega}(t,r)
= \Omega_0 \exp\left(-\frac{t^2}{2\sigma^2}-\frac{2r^2}{w^2}\right),
\qquad -\frac{\tau}{2} \le t \le \frac{\tau}{2},
\end{equation}
$\tau$ is the pulse duration, and the $\pi$ pulse duration is 62.3$\mu$s. The Gaussian beam waist $w_0$ in the transverse direction is 10 mm. In both the model and the experiment, a fixed beam radius of $w_0 = 10$ mm is used.

For the process where atomic cloud splitting, reflection, and recombine upon interaction with $\pi/2 - \pi -\pi/2 $ Bragg pulses, we employ the Runge-Kutta(RK4) method to solve the equation of motion of the atom in the Bragg pulses.
Let the atomic wave function at time $t$ be $\psi(t) = \boldsymbol{\mathit{Y}}$, then we define $f_{ge}(t, \boldsymbol{\mathit{Y}})$ as the time evolution rate
\begin{equation}
f_{ge}(t, \boldsymbol{\mathit{Y}}) = \frac{1}{i} H(t) \boldsymbol{\mathit{Y}}.
\label{a3}
\end{equation}

RK4 method updates as follows
\begin{equation}
\begin{aligned}
K_1 &= h \cdot f_{ge}(t_n, \boldsymbol{\mathit{Y}}_n) \\
K_2 &= h \cdot f_{ge} \left( t_n + \frac{h}{2}, \boldsymbol{\mathit{Y}}_n + \frac{K_1}{2} \right) \\
K_3 &= h \cdot f_{ge} \left( t_n + \frac{h}{2}, \boldsymbol{\mathit{Y}}_n + \frac{K_2}{2} \right) \\
K_4 &= h \cdot f_{ge}(t_n + h, \boldsymbol{\mathit{Y}}_n + K_3)
\end{aligned}
\end{equation}
where $h$ is the time step size, and $Y_n$ is the state vector at the current time step.

The updated state is then computed using the weighted average
\begin{equation}
\boldsymbol{\mathit{Y}}_{n+1} = \boldsymbol{\mathit{Y}}_n + \frac{1}{6} \left( K_1 + 2 K_2 + 2 K_3 + K_4 \right).
\end{equation}
The RK4 method is applied at each pulse stage to calculate the state of the atom after each pulse. After the interference is completed, the atomic cloud is freely flight for a time $t_d$ before enter detection region.
The detected light intensity is Gaussian distribution, which depends on the horizontal position of the atoms. Detection window is oriented in the north-south direction, with the $x$-axis representing the north-south axis, which can be expressed as 
\begin{equation}
W_I(r_x', v_x) = \exp \left( - \frac{2 \left[ (r_x' + v_x (t_{on} / 2))^2 + \Delta^2 \right]}{D_w^2} \right)
\end{equation}
where $\Delta$ represents the distance between the probe window and the center of the probe light in the longitudinal direction, $D_w$ is the waist of the probe light, $r_x'$ is the position of the atom when the detection light turns, and $t_{on}$ is the continue time of the probe light. The size of the detection window determines which atoms enter the detection area. Based on the window size, a cutoff is set to determine whether atoms passed through the detection window during the period when the detection light was active.

After obtaining the probability of the atomic cloud upper and lower arms reaching the final state, all atomic interference fringes undergo normalization. The phase $\phi_{cor} = 2 k_{eff}^2 \Omega_c \nu_x T^2$ of the Coriolis force effect caused by Earth's rotation is applied to the interferometer's upper arm.
\vfill 
\providecommand{\noopsort}[1]{}\providecommand{\singleletter}[1]{#1}%

\end{document}